\documentclass[letterpaper]{jpconf}

\pdfoutput=1
\usepackage{citesort}
\usepackage{subfigure}

\usepackage{graphicx}
\begin{document}
\title{Recent Results of Point Source Searches with the IceCube Neutrino Telescope}

\author{Erik Strahler for the IceCube Collaboration~\footnote{http://www.icecube.wisc.edu}}

\address{University of Wisconsin-Madison}


\begin{abstract}
IceCube is a km$^3$ scale neutrino detector being constructed deep in the Antarctic ice.  When complete, IceCube will consist of 4800 optical modules deployed on 80 strings between 1450 and 2450 m of depth.  During the 2007-2008 data taking season, 22 strings were operational.  This configuration is already much larger than previous neutrino telescopes and provides better sensitivity to point sources of high energy ($>$1 TeV) neutrino emission.  Such astrophysical objects are leading candidates for the acceleration of cosmic rays.  We describe the IceCube detector and present the methods and results of several recent searches for steady (e.g. AGN) and transient (GRB) point sources.
\end{abstract}

\section{Introduction}

High energy neutrinos provide a unqiue perspective on the cosmos.  As neutral and weakly interacting particles they point back to their sources over all energy ranges and distance scales.  This can be contrasted to cosmic rays which are bent in galactic and intergalactic magnetic fields, or photons, which can be blocked by intervening matter or attenuated at high energies via pair production with background light.  Thus neutrinos will allow us to view the deep universe with a clarity unparalled by other channels.  In addition, the detection of sources of astrophysical high energy neutrinos would be a clear signature of hadronic acceleration and a possible answer to the question of the origins of the ultra-high energy cosmic rays. Only a select few source classes, among them Gamma-ray bursts (GRBs) and active galactic nuclei (AGN), have the combination of magnetic fields and source size necessary to accelerate cosmic rays to the observed $10^{20}$ eV energies~\cite{araa:22:425}.  It is thus natural to search for neutrino emission from these sources.

In the case of hadronic acceleration, protons (or heavy nuclei) will undergo interactions with other hadrons or observed photons to produce neutrinos whose energy spectra will track that of the target field.  Given that the cosmic rays are thought to be produced with an $E^{-2}$ spectrum via the Fermi mechanism, searches for astrophysical neutrinos typically use such a power law as a benchmark.

\section{Neutrino Detection}

The IceCube detector~\cite{app:26:155} consists of an array of Digital Optical Modules (DOMs) arranged vertically on `strings' buried deep in the antarctic ice of the south pole.  Scheduled to be completed in 2011, it will consist of 4800 DOMs on 80 strings, instrumenting a cubic kilometer of clear ice at depths of 1450-1250 m.  Six additional densely instrumented strings will make up DeepCore, extending the detector sensitivity to lower energies.  A surface array, IceTop, can be used to study cosmic rays and as a particle veto.  Each DOM consists of a 25\,cm photomultiplier tube (PMT), a glass pressure sphere, and associated electronics for waveform capture and digitization~\cite{nim:a601:294}.  The detector operated with a 22-string configuration from May of 2007 through April of 2008.

Occasionally, a neutrino passing through the Earth will interact with a rock or ice nucleus near the detector, creating a charged lepton.  This lepton retains a significant portion of the primary neutrino energy, and emits Cherenkov radiation as it travels at superluminal speed through the detector.  Depending on the lepton flavor, a cascade-like or track-like light pattern will be created.  High energy neutrino-induced muons will generate tracks with a range of several kilometers.  We focus on these muons in the analyses described here as they provide good directional resolution.

The timing and amplitude of the Cherenkov photons from passing muons are recorded by the PMTs and used to reconstruct the energy and direction of the particles~\cite{nim:a524:169}.  A pointing resolution of about $1.5^{\circ}$ is possible with the 22 string configuration of IceCube.  The energy of the muon may be reconstructed to a precision of about 0.4 in $\log_{10}(\mathrm{E_{rec}}/\mathrm{E_{true}})$.  Both the directional and energy reconstructions will be leveraged for the separation of signal from background in point source searches.

\section{Point Source Searches}
\label{sec:ps}

Searches for astrophysical neutrinos face significant backgrounds.  Muons created by cosmic ray interactions in the atmosphere dwarf the signal by six to eight orders of magnitude, even at a depth of several kilometers.  From the opposite hemisphere, however, this muon background vanishes.  We therefore utilize the Earth as a filter to form a sample of muons induced by neutrinos.  The reconstruction process is not perfect though, and we must apply a series of quality criteria to remove the misreconstructed donwgoing muons from our data sample.  Even after this selection, we are left with an irreducible background of muons from cosmic ray-induced neutrinos.  In order to distinguish these from the astrophysical signal we must either search for clustering in the sky, associations with known point sources, or take advantage of the differences in energy spectrum.  Atmospheric neutrinos follow an $E^{-3.7}$ power law~\cite{pr:d70:023006}, while the signal spectrum is typically harder, depending on the details of the model.

\subsection{Time Integrated}

Two main searches for point sources of neutrinos were performed on 275.7 days of high quality data taken with the 22-string IceCube detector.  After the removal of downgoing muon backgrounds, a sample of 5114 neutrino candidates remained.  The first search looked for the most significant point on the sky, employing an unbinned maximum-likelihood analysis technique.  The signal and background are each assigned probability density functions (PDFs) describing the shapes of the distributions.  In this search, event direction and energy information were used as differentiators.  A likelihood function containing the PDFs is maximized with respect to a varying signal input for each location on the sky.  The method is described fully in~\cite{app:29:299}.  The second search used an \textit{a priori} source list of likely candidates for hadronic acceleration and performed a search using the same method as described above.  The significance of the likelihood is determined by scrambling the right ascension of the data and determing how many of these randomized datasets contain more significant observations.

\begin{figure}[t]
\centering
\includegraphics[width=0.8\textwidth]{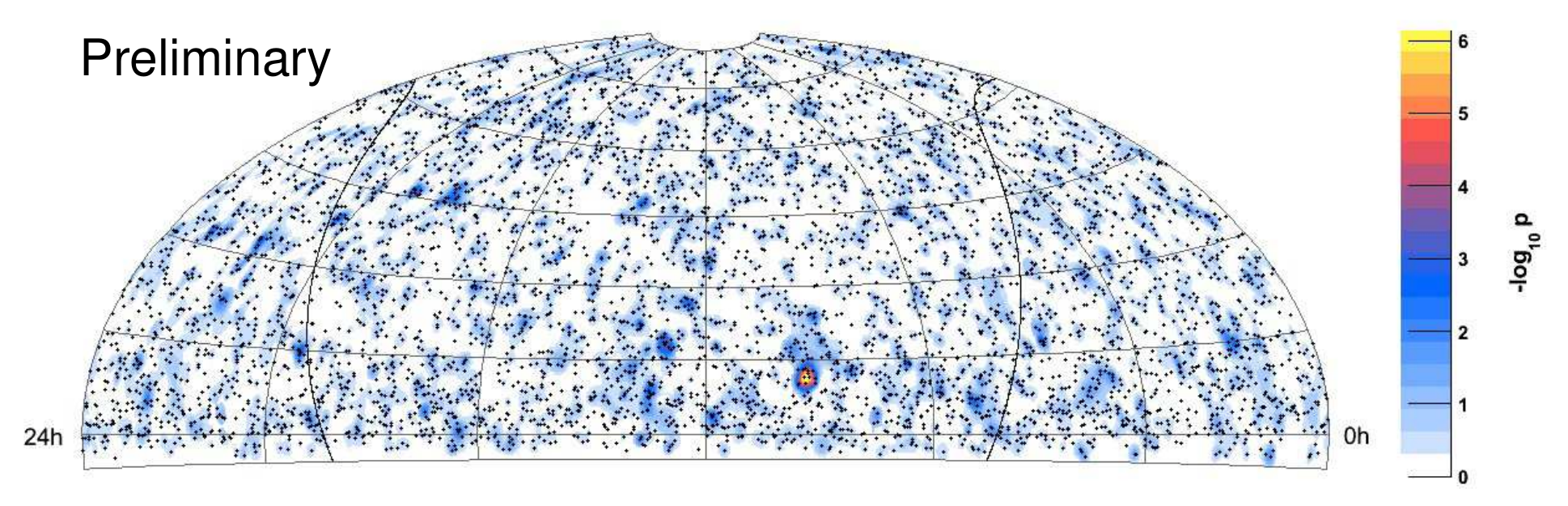}
\caption{Skymap of 5114 neutrino candidates after final event selection.  Pretrial p-values for the all-sky unbinned likelihood search are shown.}
\label{fig:skymap}
\end{figure}

When the likelihood analysis was performed on the full set of unscrambled data, the most significant location on the skymap was found to lie at $RA=153^{\circ},Dec=11^{\circ}$ (see Fig.~\ref{fig:skymap}).  The pre-trials significance was $4.8\sigma$, with a $2.2\sigma$ post-trials significance.  For the source list search, no source was found to be significant.  These results are consistent with the background only hypothesis.  Further discussion can be found in~\cite{ic22-ps}.

\subsection{Transient}

Gamma-ray bursts represent one of the plausible sources for the acceleration of the highest energy cosmic rays.~\cite{prl:75:386,prl:78:2292}.  Well developed theories of neutrino emission before~\cite{pr:d68:083001}, during~\cite{prl:78:2292}, and after~\cite{apj:541:707} the observed photon emission make them ideal candidates for searches with IceCube.  Localized in both space and time, GRBs offer excellent background rejection potential.  Furthermore, the spectral characteristics of the photon emission are generally well-measured and thus allow modelling of the expected neutrino emission.  

\begin{figure}[t]
\centering
\subfigure[prompt GRB spectra]{
\includegraphics[width=0.47\textwidth]{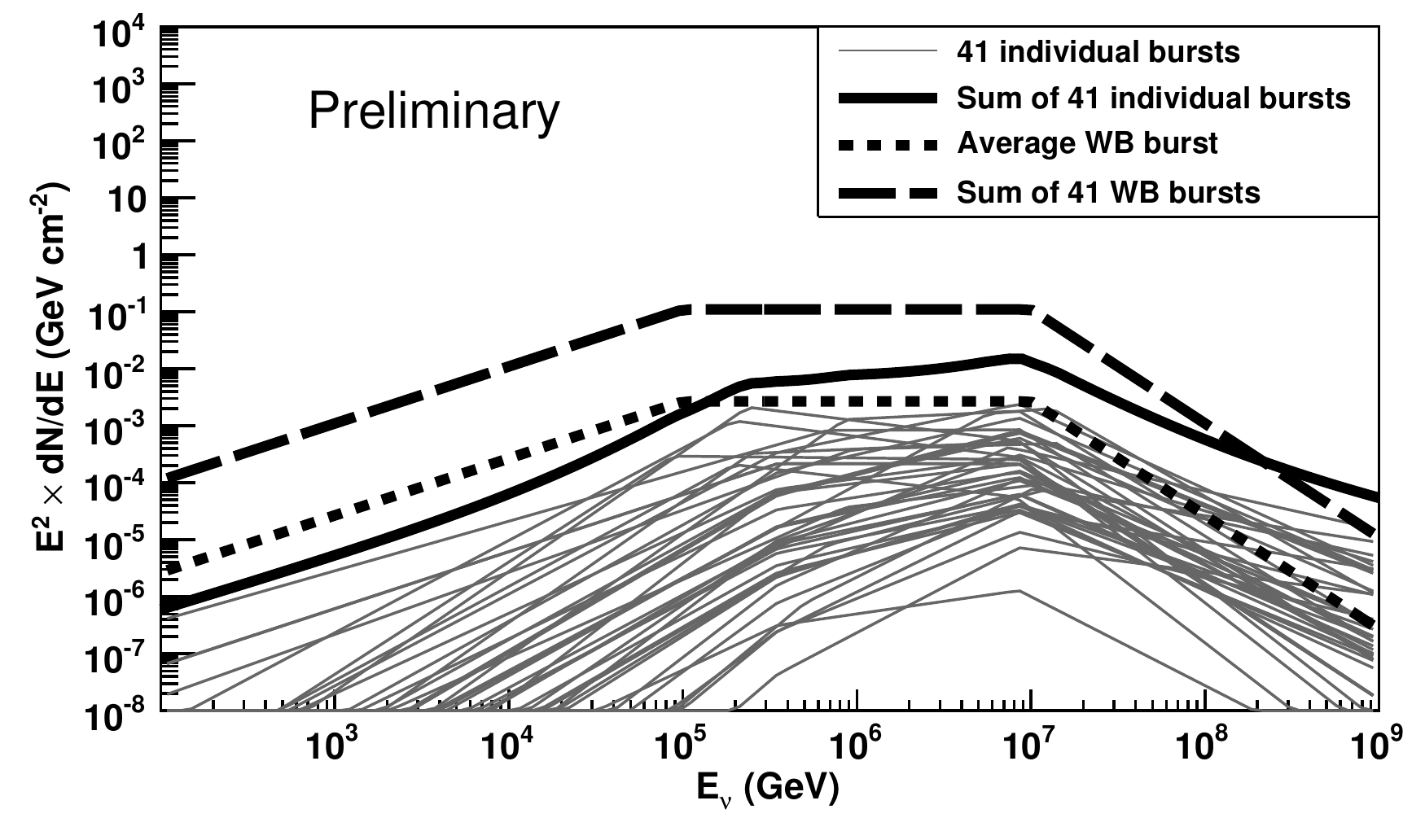}
\label{fig:ind-fluxes}
}
\subfigure[GRB search sensitivity]{
\includegraphics[width=0.47\textwidth]{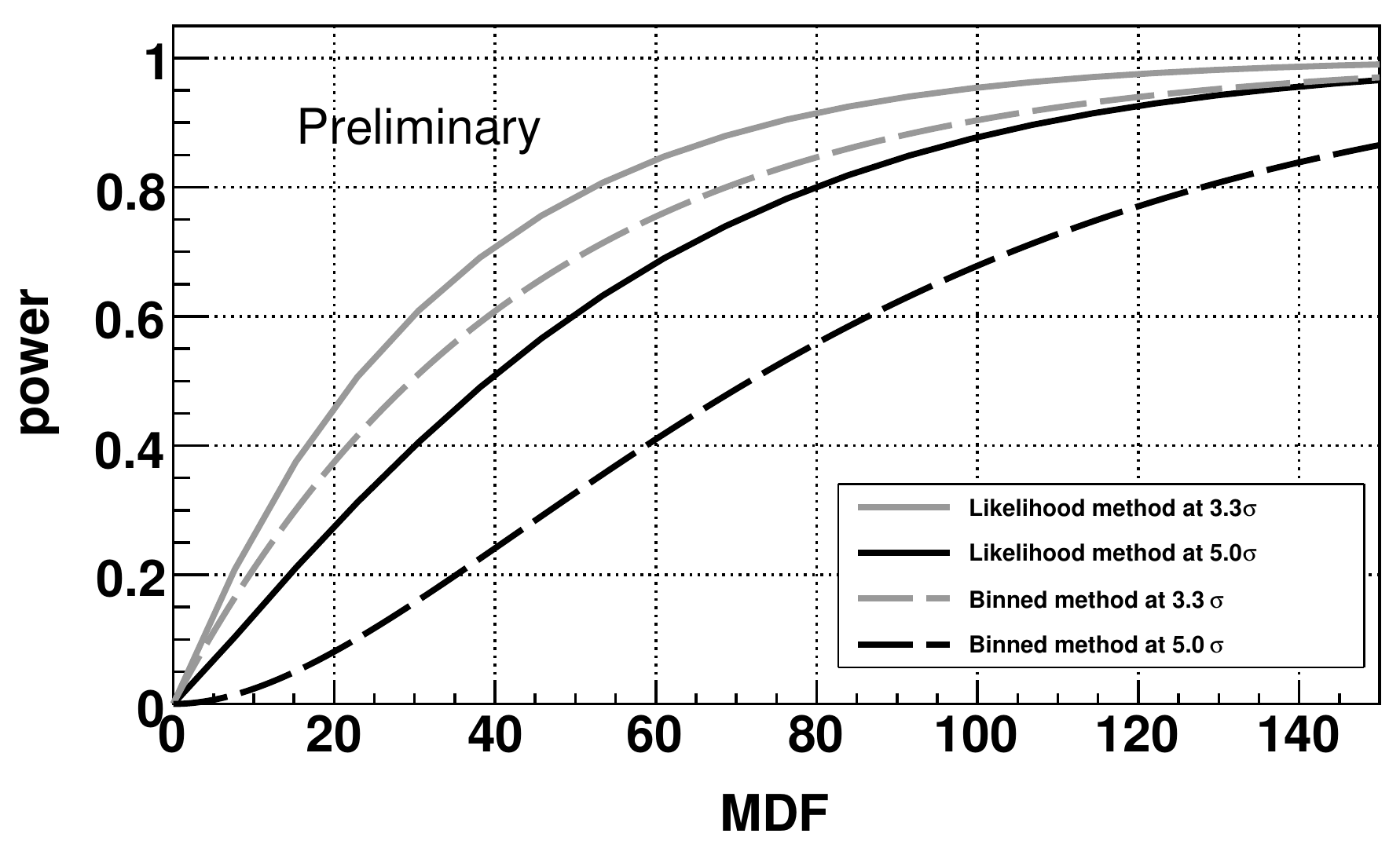}
\label{fig:sens}
}
\caption{Panel~\subref{fig:ind-fluxes} shows prompt phase neutrino spectra calculated from observed photon parameters for 41 GRBs observed in 2007-2008.  The benchmark spectrum calculated by Waxman and Bahcall using average parameters measured by the BATSE satellite is shown for reference.  Panel~\subref{fig:sens} compares the discovery potential of the unbinned search with a complementary binned search for neutrino emission during the prompt phase.  MDF indicates the multiple of the predicted emission that is required to make a discovery at the stated statistical power.  The unbinned method shows about a factor 1.7 improvement over the binned search for a 5$\sigma$ discovery at 50\% power.}
\end{figure}

For the 22-string configuration of IceCube, we have performed a stacked search for 41 GRBs in the northern hemisphere that were observed by satellites.  Because neutrino emission is predicted to occur during the \textit{precursor}, \textit{prompt}, and \textit{afterglow} phases of the emission, we perform a set of complementary searches.  For the prompt search we use the prescription of Guetta et al.~\cite{app:20:429} and the measured prompt photon spectra to model the neutrino emission from each burst (see Fig.~\ref{fig:ind-fluxes}).  Since the precursor emission is expected to occur while the GRB progenitor is still optically thick to photons, we use the prediction of Razzaque et al.~\cite{pr:d68:083001} for all bursts.  Finally, we perform a generic search for an $E^{-2}$ spectrum in an \textit{extended} window spanning \{-1,+3\} hours around each GRB.  This is expected to encompass parts of the early afterglow as well as allow for model independence in the analysis.

The GRB search started with the same event selection as described in section~\ref{sec:ps}.  A likelihood method very similar to that used in the time integrated analysis is used.  The signal PDF for each GRB consists of a spatial, temporal, and energy based term.  The PDFs for the background are taken from off-time data.  The signal PDFs from all bursts are stacked, with each contribution weighted by the expected emission from that burst, based on the individual spectral modeling and the declination of the burst.  The sensitivity of this analysis can be expressed in terms of the multiple of the predicted flux that could be excluded at 90\% C.L.  In this case, the prompt search can exclude a flux 73 times the summed signal prediction, while for the precursor search, the factor is 4.8.  When the data were unblinded, the result of the likelihood search was consistent with the background only hypothesis in all three emission windows.  Discussion of systematics and final limits from this search will be available soon and can be found in~\cite{ic22-grb}.

\section{Outlook}

While no astrophysical neutrinos have been detected to date, sensitivities are rapidly improving.  IceCube has been operating in a 40-string configuration since April 2008 and first analyses of these data are nearing maturity.  With this geometry IceCube is already as large as the full detector along its major axis, benefitting from increased angular resolution in that direction.  19 additional strings were deployed during the 2008-2009 austral summer, and the 59-string detector is scheduled to begin taking data in May of 2009.

The Fermi Gamma Space Telescope~\cite{fermi:homepage}, launched in July 2008, has nearly tripled the rate of GRB observations, increasing the opportunities for the detection of coincident neutrinos.  In additon, Fermi will detect many new steady sources of high energy photon emission.  This will translate into expanded source lists for future searches with IceCube.

The full IceCube detector will be complete in only a few years.  Initial studies of the sensitivity to high energy neutrinos from GRBs~\cite{proc:icrc09:kappes:1} are  encouraging, with preliminary results showing that IceCube will be able to detect the leading models with a high level of significance within the first year of operation.  Prospects for the detection of emission from other point sources are similarly promising.  In any case, IceCube will soon see the first sources of astrophysical neutrinos or else will set strong constraints on the emission of such neutrinos.

\section*{References}
\bibliography{myabrv,ref}

\providecommand{\newblock}{}
\begin{thebibliography}{10}
\expandafter\ifx\csname url\endcsname\relax
  \def\url#1{{\tt #1}}\fi
\expandafter\ifx\csname urlprefix\endcsname\relax\def\urlprefix{URL }\fi
\providecommand{\eprint}[2][]{\url{#2}}

\bibitem{araa:22:425}
Hillas A 1984 {\em Ann.\ Rev.\ Astron.\ Astrophys.\/} {\bf 22} 425

\bibitem{app:26:155}
Achterberg A, ({IceCube Coll}) {\em et~al.\/} 2006 {\em Astropart.\ Phys.\/}
  {\bf 26} 155 (\textit{Preprint} \eprint{astro-ph/0604450})

\bibitem{nim:a601:294}
Abbasi R, ({IceCube Coll}) {\em et~al.\/} 2009 {\em Nucl.\ Inst.\ Meth.\/} {\bf
  A601} 294 (\textit{Preprint} \eprint{arXiv:0810.4930})

\bibitem{nim:a524:169}
Ahrens J, ({AMANDA Coll}) {\em et~al.\/} 2004 {\em Nucl.\ Inst.\ Meth.\/} {\bf
  A524} 169 (\textit{Preprint} \eprint{astro-ph/0407044})

\bibitem{pr:d70:023006}
Barr G~D {\em et~al.\/} 2004 {\em Phys.\ Rev.\/} {\bf D70} 023006
  (\textit{Preprint} \eprint{astro-ph/0403630})

\bibitem{app:29:299}
Braun J {\em et~al.\/} 2008 {\em Astropart.\ Phys.\/} {\bf 29} 299
  (\textit{Preprint} \eprint{arXiv:0801.1604})

\bibitem{ic22-ps}
Abbasi R, ({IceCube Coll}) {\em et~al.\/} 2009  (\textit{Preprint}
  \eprint{hep-ph/0010323})

\bibitem{prl:75:386}
Waxman E 1995 {\em Phys.\ Rev.\ Lett.\/} {\bf 75} 386

\bibitem{prl:78:2292}
Waxman E and Bahcall J~N 1997 {\em Phys.\ Rev.\ Lett.\/} {\bf 78} 2292
  (\textit{Preprint} \eprint{astro-ph/9701231})

\bibitem{pr:d68:083001}
Razzaque S, Meszaros P and Waxman E 2003 {\em Phys.\ Rev.\/} {\bf D68} 083001
  (\textit{Preprint} \eprint{astro-ph/0303505})

\bibitem{apj:541:707}
Waxman E and Bahcall J~N 2000 {\em ApJ\/} {\bf 541} 707 (\textit{Preprint}
  \eprint{astro-ph/9909286})

\bibitem{app:20:429}
Guetta D {\em et~al.\/} 2004 {\em Astropart.\ Phys.\/} {\bf 20} 429
  (\textit{Preprint} \eprint{astro-ph/0302524})

\bibitem{ic22-grb}
Abbasi R, ({IceCube Coll}) {\em et~al.\/} 2009 In Preparation

\bibitem{fermi:homepage}
{NASA} 2008 {Fermi Gamma Space Telescope}
  \urlprefix\url{http://www.nasa.gov/mission\_pages/GLAST/main/index.html}

\bibitem{proc:icrc09:kappes:1}
Kappes A, Roth P and Strahler E 2009 {\em Proc. International Cosmic Ray
  Conference ({ICRC}'09)\/} ({\L}\'{o}d\'{z}, Poland)

\end{thebibliography}

\end{document}